\documentclass[letterpaper,conference]{IEEEtran}
\usepackage[utf8]{inputenc}
\usepackage{pdflscape}
\usepackage{afterpage}

\usepackage{cite}
\renewcommand{\thetable}{\Roman{table}}

\ifCLASSINFOpdf
  \usepackage[pdftex]{graphicx}
  \graphicspath{{images/}}
  \DeclareGraphicsExtensions{.pdf,.jpeg,.png,.eps}
\else
\fi

\usepackage{array}
\usepackage{tabularx}

\begin{document}

\title{On the Performance and Energy Efficiency of the PGAS Programming Model on Multicore Architectures}

\author{\IEEEauthorblockN{Jérémie Lagravière \&}
\IEEEauthorblockN{Johannes Langguth}
\IEEEauthorblockA{Simula Research Laboratory\\
NO-1364 Fornebu, Norway\\
\small jeremie@simula.no\\
\small langguth@simula.no
\\
}

\and
\IEEEauthorblockN{Mohammed Sourouri}
\IEEEauthorblockA{NTNU \\ Norwegian University\\of Science and Technology\\
NO-7491 Trondheim, Norway\\
\small mohammed.sourouri@iet.ntnu.no
}
\and
\IEEEauthorblockN{Phuong H. Ha}
\IEEEauthorblockA{The Arctic University of Norway\\
NO-9037 Tromsø, Norway\\
\small phuong.hoai.ha@uit.no\\
}
\and
\IEEEauthorblockN{Xing Cai}
\IEEEauthorblockA{Simula Research Laboratory\\
NO-1364 Fornebu, Norway\\
\small xingca@simula.no\\}
}


\maketitle

\begin{abstract}
Using large-scale multicore systems to get the maximum performance and energy efficiency with manageable programmability is a major challenge. The partitioned global address space (PGAS) programming model enhances programmability by providing a global address space over large-scale computing systems. However, so far the performance and energy efficiency of the PGAS model on multicore-based parallel architectures have not been investigated thoroughly.
In this paper we use a set of selected kernels from the well-known NAS Parallel Benchmarks  to evaluate the performance and energy efficiency of the UPC programming language, which is a widely used implementation of the PGAS model.
In addition, the MPI and OpenMP versions of the same parallel kernels are used for comparison with their UPC counterparts.
The investigated hardware platforms are based on 
multicore CPUs, both within a single $ 16 $-core node and across multiple nodes involving up to $1024$ physical cores.
On the multi-node platform we used the hardware measurement solution called High definition Energy Efficiency Monitoring tool in order to measure energy. On the single-node system we used the hybrid measurement solution to make an effort into understanding the observed performance differences, we  use the Intel Performance Counter Monitor to quantify in detail the communication time, cache hit/miss ratio and memory usage.
Our experiments show that UPC is competitive with OpenMP and MPI on single and multiple nodes, with respect to both the performance and energy efficiency.
\end{abstract}

\section{Introduction \& Motivation}
\label{intro}
The overarching complexity of parallel programming is one of the fundamental challenges that the HPC research community faces.
In the last decade, the Partitioned Global Address Space model (PGAS) has been established as one possible solution for this problem. It promises improved  programmability while maintaining high performance, which is the primary goal in HPC. In recent years, energy efficiency has become an additional goal. Optimizing energy efficiency without sacrificing  computational performance is the key challenge of energy-aware HPC, and an absolute requirement for attaining Exascale computing in the future. 

In this study we investigate whether PGAS can meet the goals of performance and energy efficiency. We focus on UPC, one of the most widely used PGAS implementations, and compare it to MPI and OpenMP.
OpenMP offers ease of programming for shared memory machines, while MPI offers high performance on distributed memory supercomputers.

PGAS combines these advantages through a simple and unified memory model. On a supercomputer, this means that the programmer can access the entire memory space as if it is a single memory space that encompasses all the nodes. Through a set of functions that makes data \textsl{private} or \textsl{shared}, PGAS languages ensure data consistency across the different memory regions. When necessary, shared data is transferred automatically between the nodes through a communication library such as GASnet \cite{gasnetOfficial}.  

Recent studies \cite{advocatePGAS2,advocatePGAS1} advocate the use of PGAS as a promising solution for HPC. Many have focused on the evaluation of PGAS performance and UPC in particular \cite{upcPerformanceStudy2002,upcBenchmarkingIssues2001,upcPerformance2003,UPC_MVAPICH_NAS,upcPerformance2009,shan2010programming,upcPerformande2007}. However, the previous UPC studies have not taken energy efficiency into consideration. This motivates us to investigate UPC's energy efficiency and performance using the latest CPU architecture with advanced support for energy and performance profiling.

For our evaluation we use the well-established NAS Benchmark. We use MPI, OpenMP, and UPC implementations \cite{NAS_Official,UPCNAS} to compare the performance and energy efficiency of the different programming models. The energy measurements of the single-node system are obtained by using Intel PCM \cite{IntelPCM}. The multi-node performance measurements are obtained on an Intel Xeon based supercomputer , the energy measurement are obtained on this platform by using High Definition Energy Efficiency Monitoring (HDEEM) \cite{hdeem}. We provide an analysis of the single-node measurements in order to explain the difference in performance and energy efficiency, by focusing on the cache performance and the memory traffic of MPI, OpenMP and UPC.

This paper improves upon previous works \cite{upcPerformanceStudy2002,upcBenchmarkingIssues2001,upcPerformance2003,upcPerformance2009,shan2010programming,upcPerformande2007} by:
(1) providing measurements for a larger number of nodes and cores for MPI, OpenMP and UPC on recent single-node and multi-node systems (up to 1024 physical cores); (2) including energy measurements obtained on both a single-node system and a multi-node system; (3) making an effort to understand the differences in energy efficiency and performance between UPC, OpenMP and MPI.

The remainder of this paper is organized as follows: Section \ref{pgas} briefly presents the UPC framework and why we have chosen this programming language. Section \ref{nas} describes the benchmark chosen for this study.  Section \ref{experimentalSetup} explains the hardware and software set-up used for running our experiments, the results of which are presented in Section \ref{results} and discussed in Section \ref{discussion}.
Section \ref{conclusion} concludes the paper.

\vspace{-0.15cm}
\section{PGAS Paradigm and UPC}
\label{pgas}
PGAS is a parallel programming model that has a logically partitioned global memory address space, where a portion of it is local to each process or thread. A special feature of PGAS is that the portions of the shared memory space may have an affinity for a particular process, thereby exploiting locality of reference \cite{UPCMPIOPENMPComparison2005,wikipediaDefPGAS}.

In the PGAS model, each node has access to both private and shared memory. Accessing the shared memory to either read or write data can imply inter-node communication  which is handled automatically by the runtime. Remote access to memory work in an  RDMA (Remote Direct Memory Access) fashion, using one-sided communication. However, most PGAS languages are built over a low-level communication layer which limits their physical capabilities. Thus, RDMA is available only if the underlying hardware and software support it.  

In recent years several languages implementing the PGAS model have been proposed. UPC, which is essentially an extension of the C language, was one of the first ones and also one of the most stable \cite{PGAS_Definition}. Other members of the PGAS family of languages include the Fortran counterpart, Coarray Fortran \cite{coArrayFortran}, X10 \cite{x10}, and Cray Chapel \cite{crayChapel}. In addition, libraries such as Global Arrays \cite{globalArray} and SHMEM/OpenSHMEM \cite{openshmem} which implement PGAS functionality are available.

The key characteristics of UPC are: a parallel execution model of Single Program Multiple Data (SPMD); distributed data structures with a global addressing scheme, with static or dynamic allocation; operators on these structures, with affinity control; and copy operators between private, local shared, and distant shared memories.

Additionally, multiple open-source implementations of the UPC compiler and runtime environment are available, in particular Berkeley UPC \cite{BerkeleyUPC}, GCC/UPC \cite{GCCUPC} and CLANG/UPC\cite{clangUPC}.

\section{The NAS Benchmark}
\label{nas}
The NAS Benchmark \cite{NASOrigin} consists of a set of kernels that each provides a different way of testing the capabilities of a supercomputer. The NAS Benchmark was originally implemented in Fortran and C. We use both the Fortran and C implementations for OpenMP and MPI, as well as the UPC version of the benchmark \cite{UPCNAS}. For our study, we select four kernels: Integer Sort (IS), Conjugate Gradient (CG), Multi-Grid (MG), and Fourier Transformation (FT).

CG refers to the \textsl{conjugate gradient} method used to compute an approximation to the smallest eigenvalue of a large, sparse, symmetric positive definite matrix. 

MG is a simplified \textsl{multigrid} kernel. Multigrid (MG) methods in numerical analysis solve differential equations using a hierarchy of discretizations. 

FT is a three-dimensional partial differential equation solver using \textsl{Fast Fourier Transformations}. This kernel performs the essence of many spectral codes. It is a rigorous test of all-to-all communication performance. 

IS represents a large \textsl{integer sort}. This kernel performs a sorting operation that is important in {particle method codes. It evaluates both integer computation speed and communication performance. 

CG, IS, MG and FT  are selected  because they are the most relevant ones: stressing memory, communication and computation. They involve very different communication patterns, which is important for evaluating the performance of the selected languages (see Section \ref{results}). The other kernels in the NAS Benchmark are of limited relevance to this study. See \cite{NASOrigin} for their descriptions.

\section{Experimental Setup}
\label{experimentalSetup}
In this section we describe the software and hardware solutions that we used to carry out our experiments. We ran the NAS kernels both on a single-node machine and on Taurus \cite{taurusOfficial}, a supercomputer operated by Dresden University of Technology, using a varying number of cores and nodes.

\vspace{-0.1cm}
\subsection{Hardware}
Table \ref{hardwareSetup} shows the specifications of the systems used in our experiments. The single-node system is equipped with Intel Sandy Bridge processors and the multi-node supercomputer is equipped with Intel Haswell processors.

\begin{table}[h]
\centering
\caption{Experimental setup: hardware}
\label{hardwareSetup}
\scalebox{0.7}{
\begin{tabular}{|l|c|c|}
\hline
\textbf{}     & \textbf{Single Node}                  & \textbf{Multi Node }                 \\ \hline
\textbf{CPUs}         & 2                            & 2   (per node)                \\ \hline
\textbf{Cores}          & 16                           & 24  (per node)              \\ \hline
\textbf{CPU model}                & Intel Xeon E5-2650 @ 2.00GHz &  Intel Xeon E5-2680 v3 @ 2.50GHz \\ \hline
\textbf{Interconnect}             & N/A                           & Infiniband: 6.8 GB/s        \\ \hline
\end{tabular}
}
\end{table}
\vspace{-0.1cm}
\subsection{Software}
On the single-node machine we used UPC version 2.22.0, the Intel Compiler version 15.0.1, and MPI Library 5.0 Update 2 for Linux. 
On the multi-node supercomputer we used UPC version 2.22.0 and using Bull XMPI version 1.2.8.4.
\subsubsection{UPC}
On the multi-node supercomputer, the UPC compiler and runtime were built with a specific option enable-segment-large in order to support large memory systems.
The UPC applications were compiled using a fix number of threads and a fix network choice:
\textsl{upcc -T1024 -network=mxm}.
The UPC implementations were run using a fixed number of threads and fixed shared heap size and thread binding:
\textsl{upcrun -n number\_of\_processes -bind-threads -shared- heap=3765MB ./application}.
On the single-node system the UPC application were using the symmetric multiprocessing (smp) network conduit.
The following environment variables were defined for \textsl{UPC GASNET\_PHYSMEM\_MAX=63G} which indicates to GASNET \cite{gasnetOfficial} to use 63GB of RAM on each node. We also used the environment variable \textsl{GASNET\_PHYSMEM\_NOPROBE=1} which indicates to GASNET to avoid memory detection on each node. Except for FT-D-16 where we used a modified GASNET environment variable: \textsl{GASNET\_PHYSMEM\_MAX=254G} because FT requires more RAM than the other kernels.
\subsubsection{MPI}
On the multi-node supercomputer and single-node system MPI applications were compiled using the \textsl{-O3 -mcmodel=medium} flag in order to handle larger data in memory.
\subsubsection{OpenMP}
On the single-node system we used OpenMP version 4.0 and we used \textsl{numactl} and the \textsl{OMP\_NUM\_THREADS } environment variable to bind the threads and define the number of threads.
\subsubsection{Benchmarking}
For each measurement we executed three separate runs and reported the best result. Doing so filters out the OS interference.

On the single-node machine we used size Class \textbf{C} \cite{NAS_Official,justifiySizeC2} for each kernel. For CG, Class C, the number of rows is $150000$. For MG, Class C, the grid size is $512 \times 512 \times 512$. For FT, Class C, the grid size is $512 \times 512 \times 512$. For IS, Class C, the number of keys is $2^{27}$. On the multi-node supercomputer we used for each kernel Class \textbf{D}, except for IS which is not available in Class \textbf{D} in the UPC implementation \cite{UPCNAS}. For CG, Class D, the number of rows is $1500000$. For MG, Class D, the grid size is $1024 \times 1024 \times 1024$. For FT, Class D, the grid size is $2048 \times 1024 \times 1024$.

In our study the comparison between the different implementations is fair as the number of operations (expressed in Million Of Operation - MOP) is identical in all implementations (OpenMP / MPI / UPC).
The number of MOP is reported by the benchmarks.
For example, the number of MOP for CG in size C is precisely 143300 for all implementations (OpenMP, MPI and UPC).

Each kernel was run using up to 1024  CPU cores and thus 64 nodes of the supercomputer. However, for CG,  limitations in the UPC implementation prevent us from using more than 256 cores, see Figure \ref{fig:cgAbel}.

Sizes \textbf{C} and \textbf{D} provide data sets that are sufficiently large to exceed the cache size of the test systems \cite{justifiySizeC1} \cite{justifiySizeC2}.

\subsubsection{Thread Binding}
Thread binding or thread pinning is an approach that associates each thread with a specific processing element.  In our experiments we applied thread/process binding to the physical cores.

\vspace{-0.1cm}
\subsection{Energy Measurements}
\label{energyMeasurement}

\subsubsection{Multi-Node Platform}
On the multinode platform we have chosen High Definition Energy Efficiency Monitoring (HDEEM) \cite{hdeem}. HDEEM is a hardware based solution for energy measurements, meaning that additional hardware is used to measure energy in a supercomputer. HDEEM is an intra-node measurement tool, which indicates that the additional hardware that performs the energy measurements is located inside each node of the supercomputer \cite{hdeem3}. The authors in \cite{hdeem} define four criteria, including spatial and temporal granularity, accuracy
and scalability, for power and energy measurement. HDEEM can achieve an accuracy of 99.5\% over 270 nodes by using an appropriate filtering approach to prevent the aliasing effect. HDEEM is based on an FPGA solution to achieve spatial fine-granularity by measuring every blade, CPU and DRAM power separately, with a sampling rate of 1,000 Sa/s over 500 nodes\cite{hdeem2}.
Our measurements do not take into account the energy consumption of the network between the nodes of the multi-nodes platform.

\subsubsection{Single-Node Platform}
On the single-node platform we have chosen a software based solution in order to measure the CPU and RAM energy consumption and memory and cache usage. Intel Performance Monitor (Intel PCM) is used for the energy efficiency experiments on the single-node platform and to measure memory and cache usage\cite{IntelPCM}.
Intel PCM uses the Machine Specific Registers (MSR) and RAPL counters to disclose the energy consumption details of an application \cite{useOfIntelPCM2}. Intel PCM is able to identify the energy consumption of the CPU(s) and the RAM. Quick-Path Interconnect energy consumption is not taken into account  because Intel PCM was unable to provide measurement on the chosen hardware platform.

The RAPL values do not result from physical measurement. They are based on a modeling approach that uses a “set of architectural events from each core, the processor graphics, and I/O, and combines them with energy weights to predict the package’s active power consumption” \cite{IntelPCMReviewer1_2}. Previous studies have demonstrated that using a counter-based model is reasonably accurate \cite{IntelPCMReviewer1_AccurateEnergyModeling1,IntelPCMReviewer1_AccurateEnergyModeling2,IntelPCMReviewer1_AccurateEnergyModeling3,IntelPCMReviewer1_AccurateEnergyModeling4}. The RAPL interface returns energy data. There is no timestamp attached to the individual updates of the RAPL registers, and no assumptions besides the average update interval can be made regarding this timing. Therefore, no deduction of the power consumption is possible other than averaging over a fairly large number of updates. For example, averaging over only 10 ms would result in an unacceptable inaccuracy of at least 10\% due to the fact that either 9, 10, or 11 updates may have occurred during this short time period~\cite{IntelPCMReviewer1}. In our experiments the execution time of each kernel is always above a second, thus estimating the power data by averaging reported Joules over time is accurate. For convenience, in our study, we use the word "measurement" to mention the values reported by Intel PCM.

We do not track and report the energy consumption curve along each kernel's execution time line, instead we choose to report the total energy consumption ($Joules$) and the average power consumption ($Watt = Joules/seconds$).

\section{Results}
\label{results}

\begin{figure}[t]
	\includegraphics[width=0.5\textwidth]{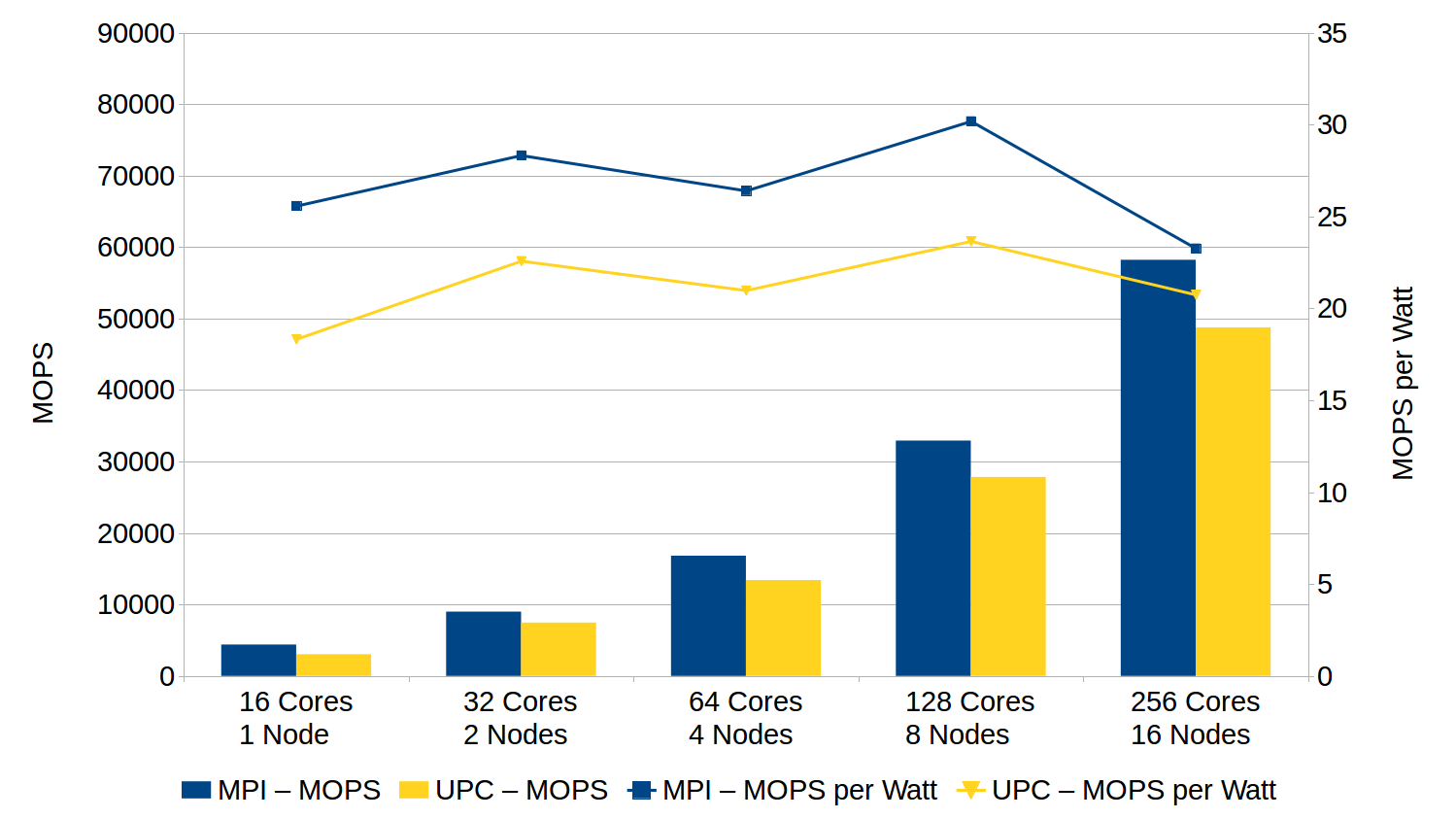}
	\caption{Multi-node performance  and energy efficiency of the CG kernel - Class D}
	\label{fig:cgAbel}       
\end{figure}
\begin{figure}[t]
	\includegraphics[width=0.5\textwidth]{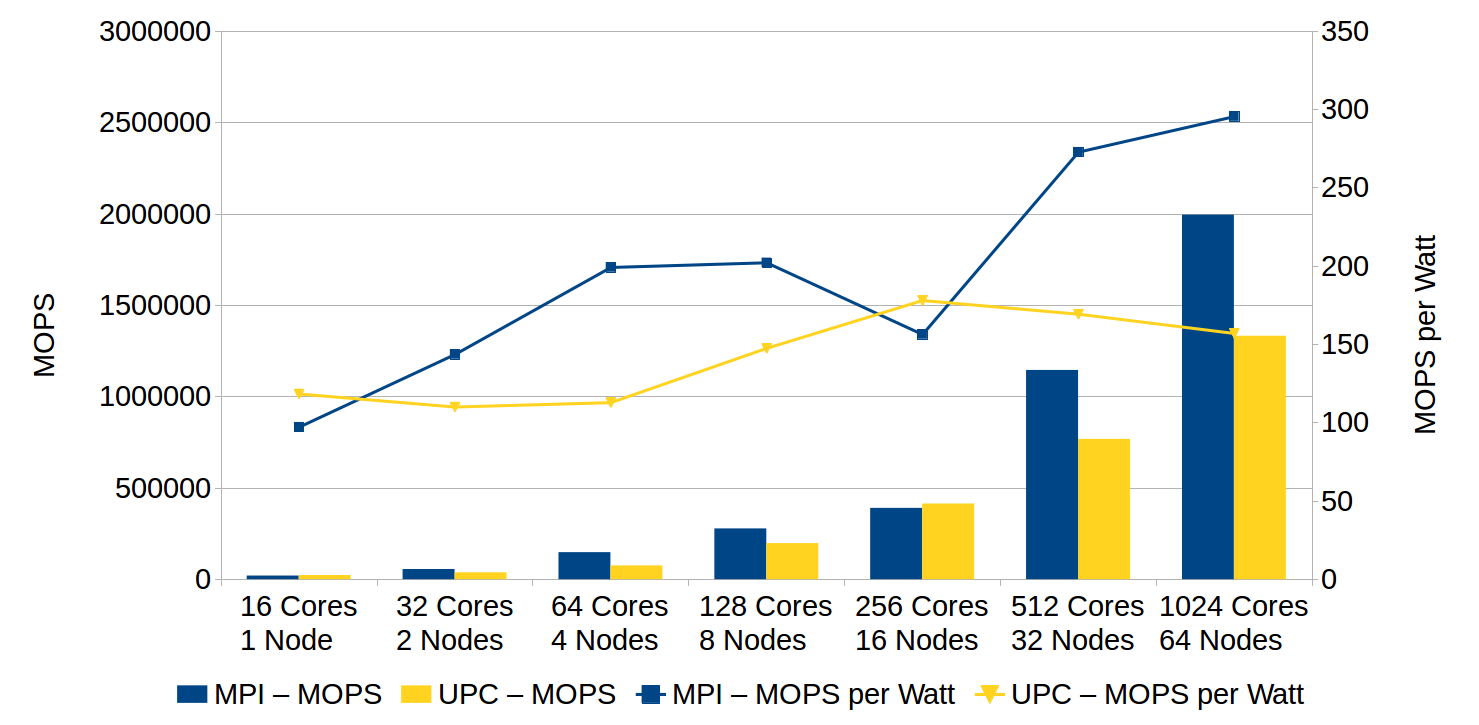}
	\caption{Multi-node performance and energy efficiency of the MG kernel - Class D}
	\label{fig:mgAbel}       
\end{figure}
\begin{figure}[t]
	\includegraphics[width=0.5\textwidth]{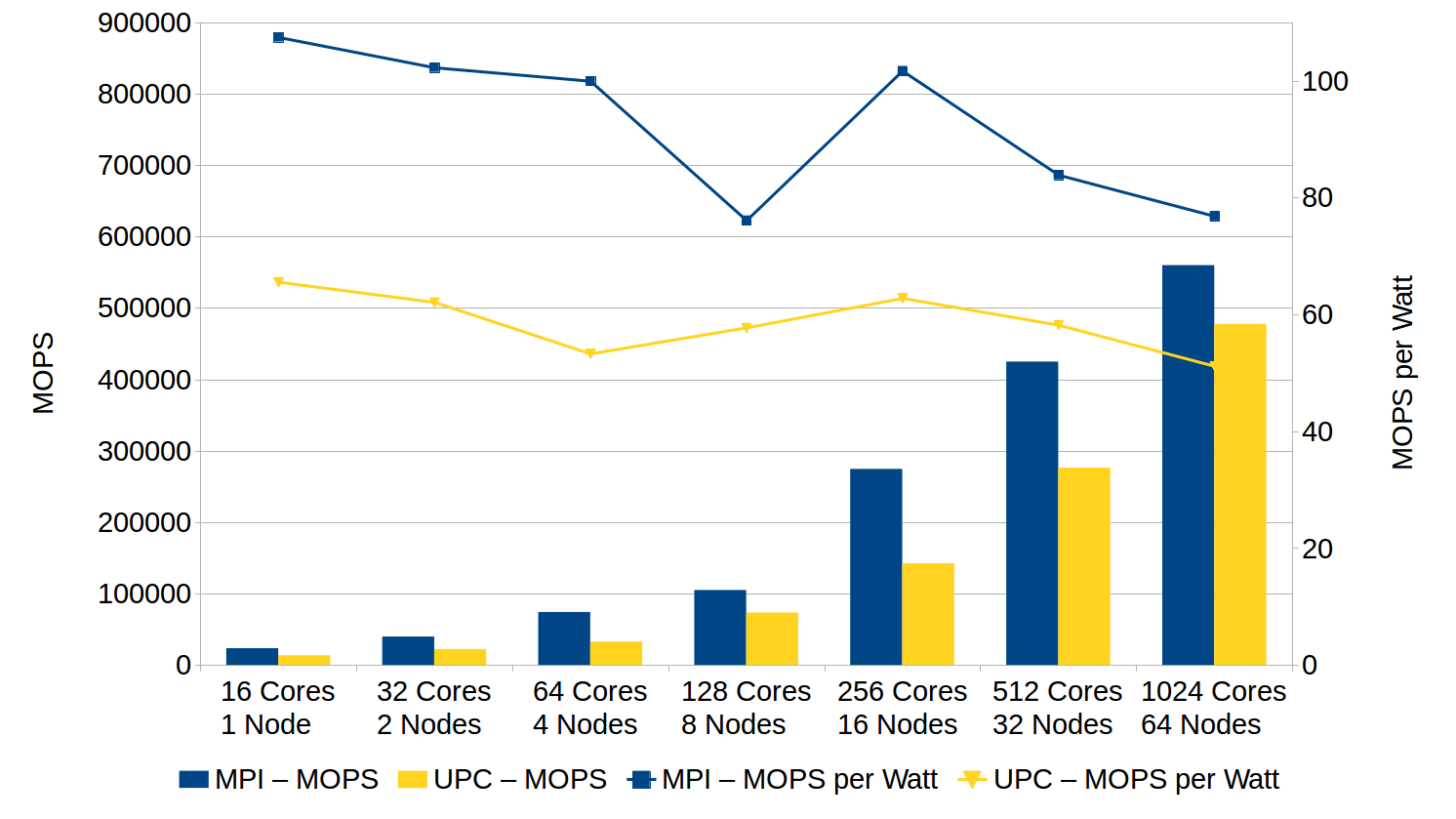}
	\caption{Multi-node performance and energy efficiency of the FT kernel - Class D}
	\label{fig:ftAbel}       
\end{figure}
\begin{figure}[t]
	\includegraphics[width=0.5\textwidth]{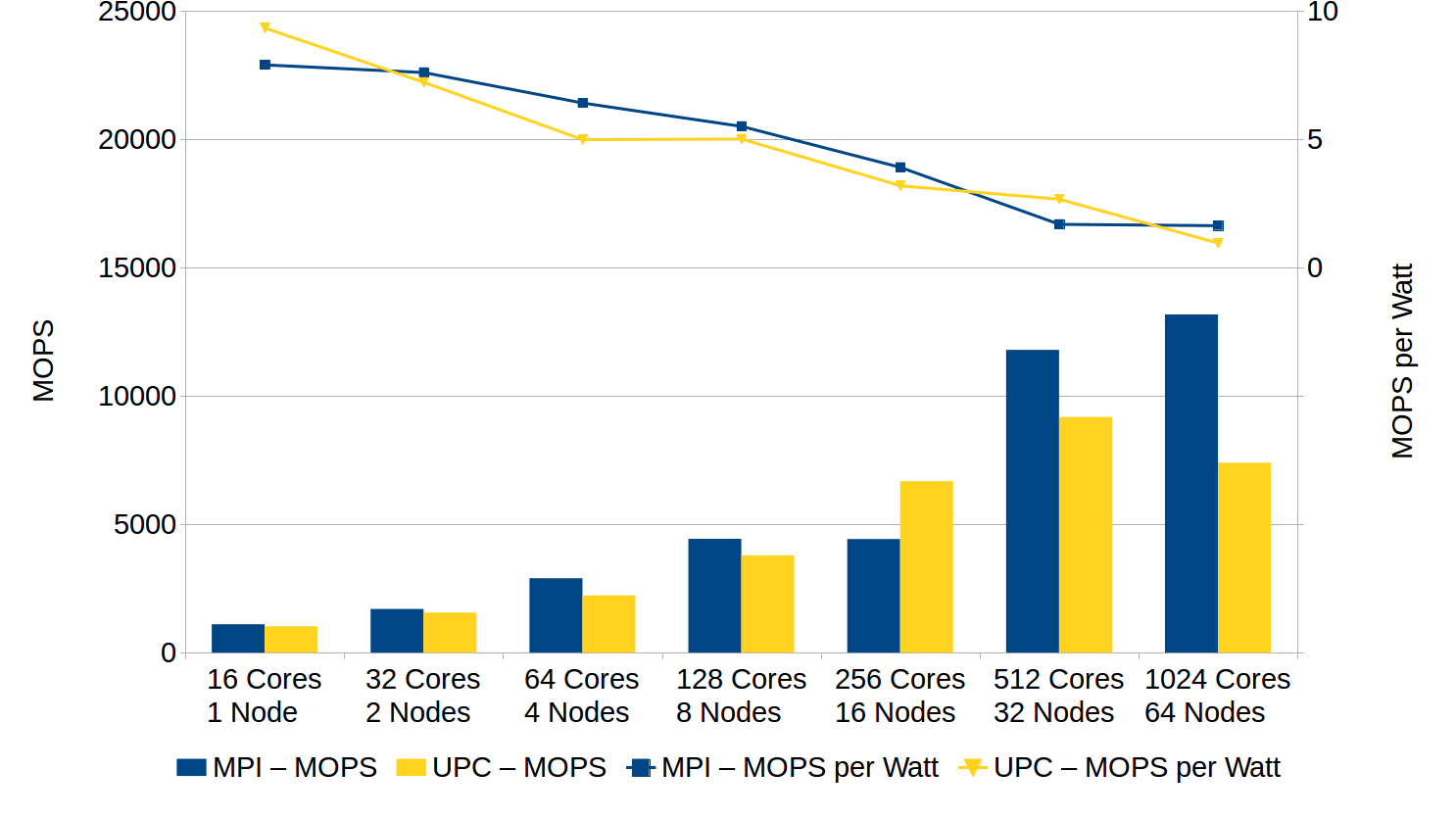}
	\caption{Multi-node performance and energy efficiency of the IS kernel - Class C. On the left Y-axis the scale has been shifted for readability purposes.}
	\label{fig:isAbel}       
\end{figure}

In this paper, we consider two metrics to measure the  performance and energy efficiency.
To evaluate the  performance we use Million Operations Per Second (MOPS). This metric is used for both the multi-node measurements and single-node measurements.
To evaluate the  energy efficiency, Millions Operations Per Seconds over Watts (MOPS per Watt) is used as the energy efficiency metric. This metric is used for both the multi-node measurements and single-node measurements. The \textsl{500 Green - Energy Efficient High Performance Computing Power Measurement Methodology} \cite{Top500Guidelines} advises this measurement methodology. We remark that MOPS per Watt is equivalent to MOP/Joules:
$(MOP / seconds)/Watt = (MOP / seconds) / (Joules / seconds) = MOP/Joules $

For single-node measurements, we use the following notation: [kernel name]-[number of threads/processes]-[1S / 2S]. For instance, \textsl{CG-8-2S} stands for "Conjugate Gradient kernel running on 8 threads (or processes for MPI) spread over two sockets".

For multi-node measurements, we use the following notation [kernel name]-[class]-[number of threads/processes]. For instance, MG-D-256 stands for the "Multigrid kernel of class D running on 256 threads (or processes for MPI)". Each node always used the maximum number of physical cores, \textsl{i.e.} 16.

\vspace{-0.15cm}
\subsection{Measurement on Multi-Node Architecture}
To the best of our knowledge, this is the first investigation of UPC's energy efficiency. Our experimental results show that the energy efficiency of UPC, MPI, and OpenMP implementations scale over the number of cores and are comparable to each other. Figures~\ref{fig:cgAbel}-\ref{fig:isAbel} show the multi-node performance expressed in MOPS and the energy efficiency expressed in MOPS per Watt, for the four kernels implemented in UPC and MPI. Each kernel ran on up to 1024 cores, except CG where the UPC implementation cannot run on more than 256 threads.

The performance results (bars) show that the CG, MG and FT kernels scale over the number of cores independently of the language. MPI is a clear winner when running on more than 32 cores, however UPC achieves a performance that is close to that of MPI, particularly for the CG and MG kernels. These results match previous studies, in particular \cite{shan2010programming}.

IS is aside in terms of performance, because of the size of the data to process, size C, is smaller than size D and causes both UPC and MPI not being able to scale on more than 256 cores. For 512 and 1024 cores runs, IS-C does not deliver good performance as the communication cost outbalances the computation performance.

Figures~\ref{fig:cgAbel}-\ref{fig:isAbel} show a complex relation between performance (bars) and energy efficiency (lines). While the performance of both UPC and MPI go up with increasing numbers of cores and nodes, the energy efficiency is at best staying constant or diminishing. The only exception is the MG benchmark, where higher energy efficiency is achieved by using more cores and nodes.
For the IS benchmark that uses the size of Class C, in particular, the energy measurements obtained by HDEEM are gradually dominated by the non-scalable initialization phase as the number cores increases. (In comparison, the energy measurements obtained by Intel PCM on the single-node system do not include the initialization phase.)

\vspace{-0.1cm}
\subsection{Measurements on Single-Node Architecture}
\begin{figure}[t]
	\includegraphics[width=0.5\textwidth]{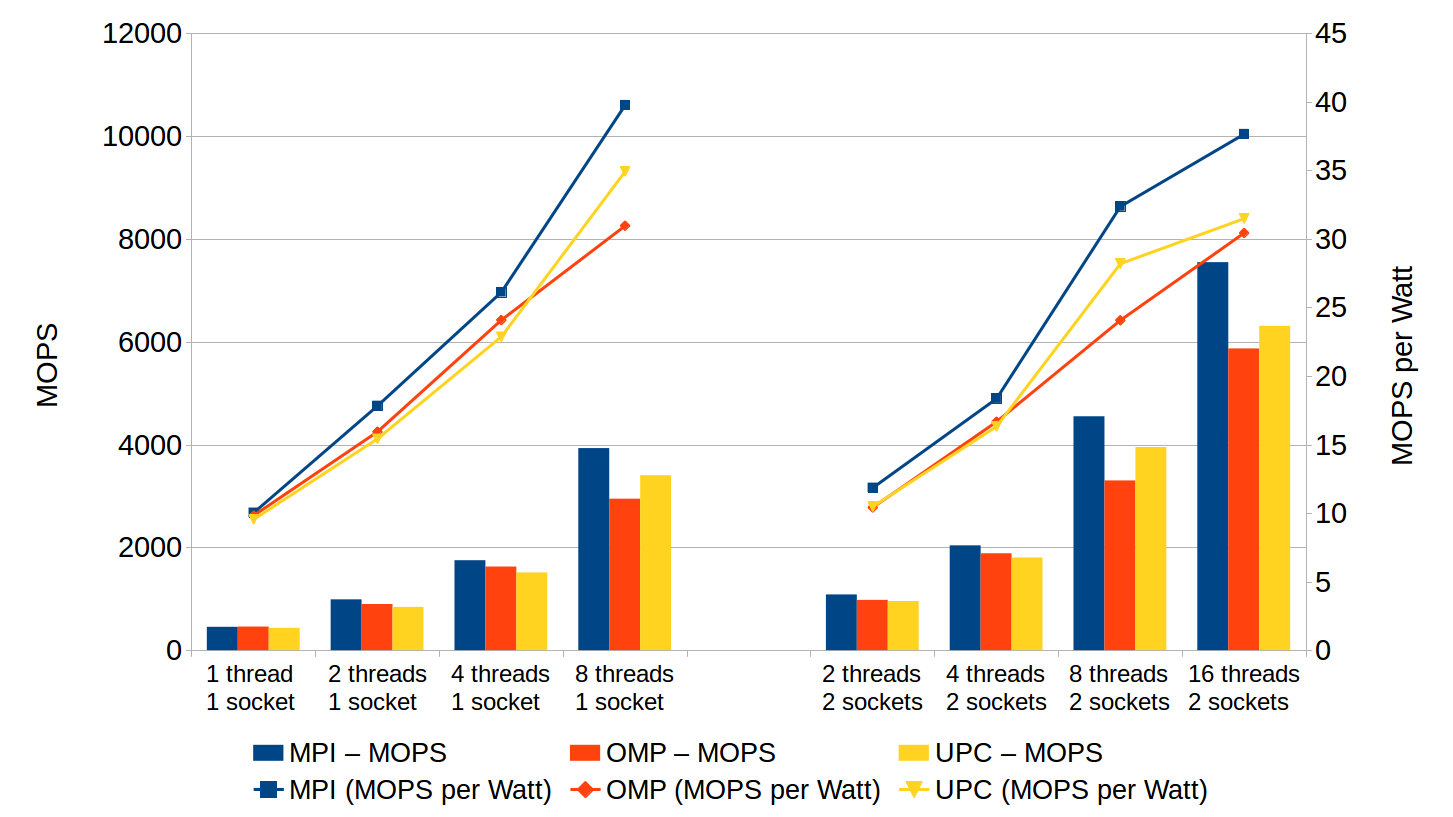}
	\caption{Single-node performance and energy efficiency of the CG kernel - Class C}
	\label{fig:cgMops}       
\end{figure}
\begin{figure}[t]
	\includegraphics[width=0.5\textwidth]{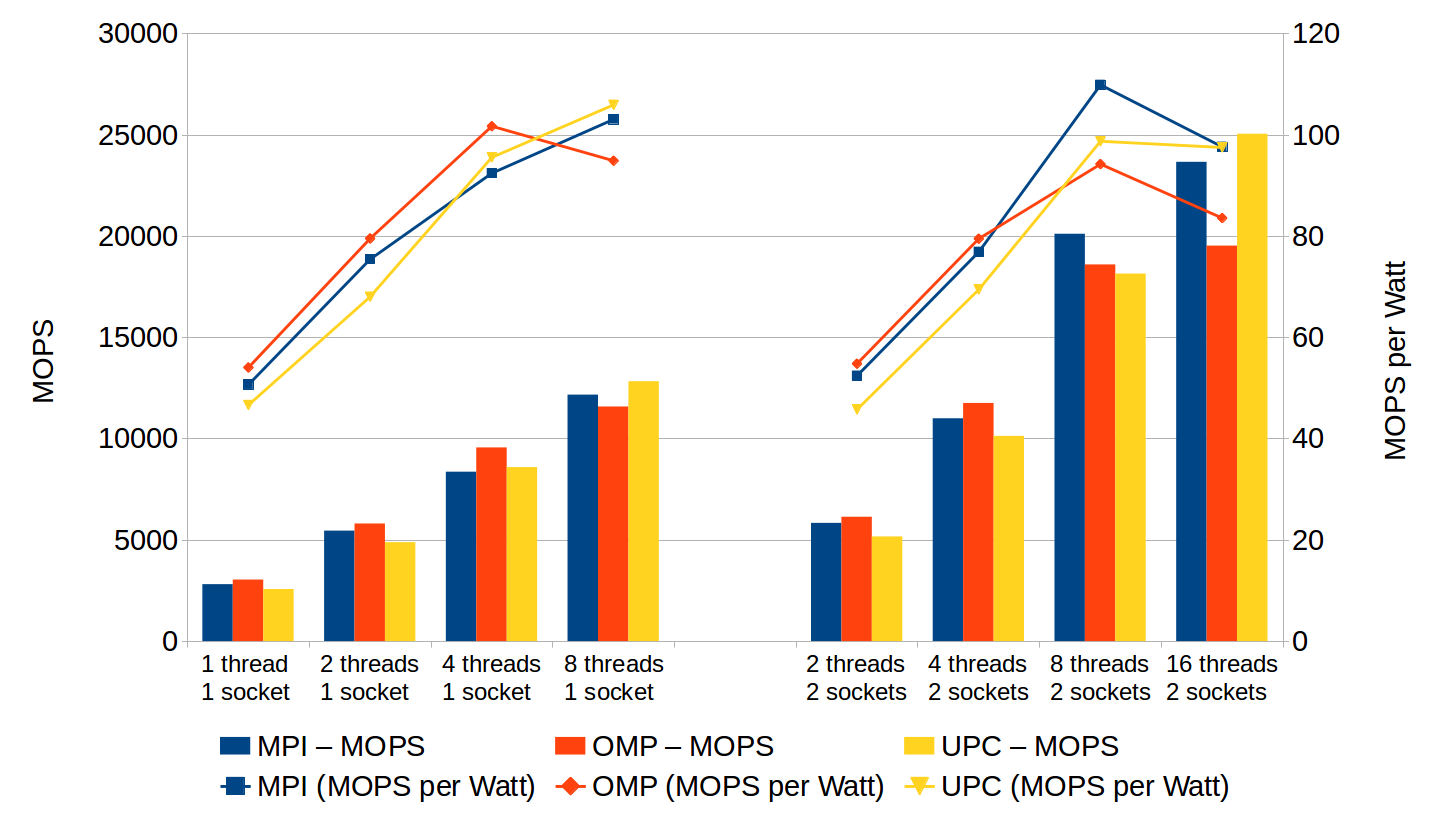}
	\caption{Single-node performance  and energy efficiency of the MG kernel - Class C}
	\label{fig:mgMops}       
\end{figure}
\begin{figure}[t]
	\includegraphics[width=0.5\textwidth]{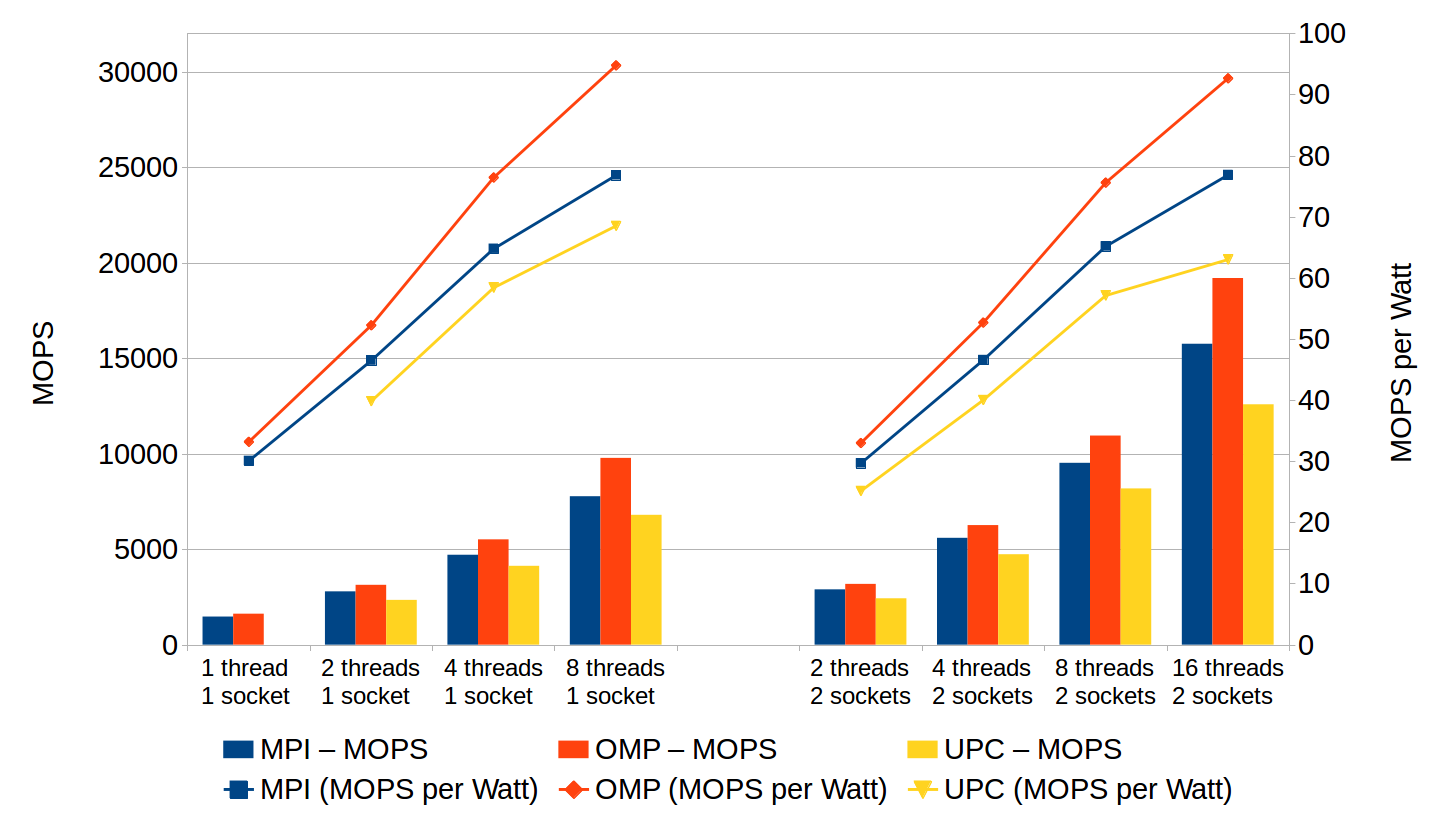}
	\caption{Single-node performance  and energy efficiency of the FT kernel - Class C}
	\label{fig:ftMops}       
\end{figure}

\begin{figure}[t]
	\includegraphics[width=0.5\textwidth]{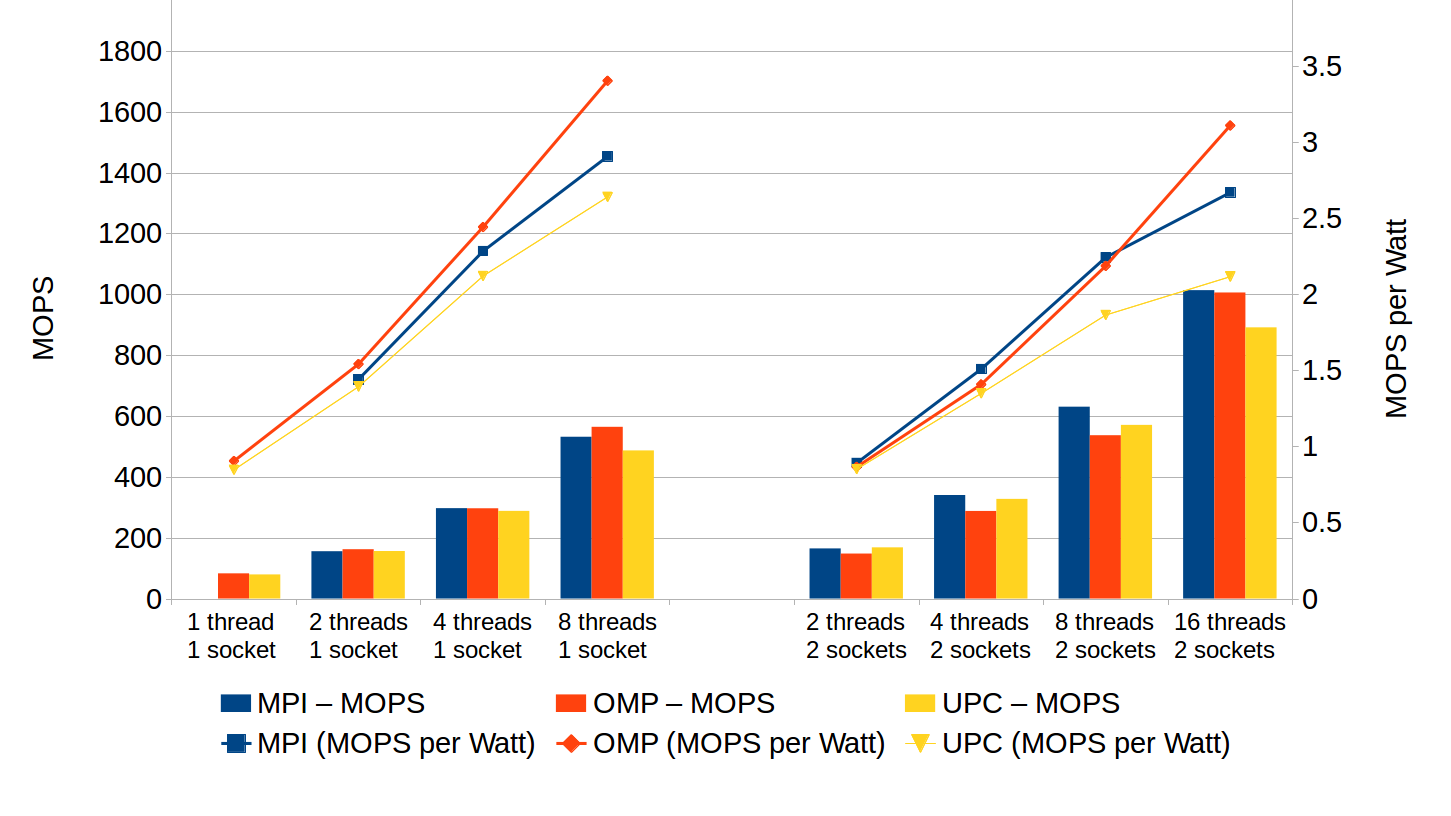}
	\caption{Single-node performance  and energy efficiency  of the IS kernel - Class C}
	\label{fig:isMops}       
\end{figure}
Our experimental results show that the energy efficiency of UPC, MPI, and OpenMP implementations scale over the number of cores and are comparable to each other. In this section we give more details about these results and in Section \ref{discussion} we provide an analysis of the difference in performance between UPC, OpenMP, and MPI.

Figures~\ref{fig:cgMops}-\ref{fig:isMops} show both the performance expressed in MOPS for the four kernels (bars) and the energy efficiency expressed in MOPS per Watt (lines). Each kernel ran on up to 16 cores. As the single-node system is equipped with two CPUs we ran the kernels on both one socket and two sockets for 2,4 and 8 threads counts. Each of these figure shows the results for all three programming models. The energy efficiency results show that the kernels scale over the number of threads/cores independently of the language. In Figures~\ref{fig:cgMops}-\ref{fig:isMops}, when measured on one-socket, the energy consumption of the idle socket and memory controller were not taken into account. There is no clear winner since none of the chosen languages is better than the other two competitors for all the kernels.

UPC was not able to run FT-C on one thread and MPI was not able run IS-C on one thread.

\begin{table*}[]
\centering
\caption{Comparison of computation and energy performance of CG over 2,4, and 8 threads using 1 and 2 sockets.\newline \tiny Syntax: \textsl{CG-2-2S / CG-2-1S} stands for "Measurements from CG running over 2 threads using 2 sockets compared to measurements from CG running over 2 threads using 1 socket" }
\label{table:energyTable}
\scalebox{0.80}{
\begin{tabular}{lrrrrrrrrr}
\cline{2-10}
\multicolumn{1}{l|}{\textbf{\begin{tabular}[c]{@{}l@{}}Second socket\\ not included\\ when not used\end{tabular}}} & \multicolumn{3}{c|}{MPI}                                                                                                                                                                                                                                                                & \multicolumn{3}{c|}{OpenMP}                                                                                                                                                                                                                                                            & \multicolumn{3}{c|}{UPC}                                                                                                                                                                                                                                                               \\ \hline
\multicolumn{1}{|c|}{CG}                                                                                           & \multicolumn{1}{c|}{\begin{tabular}[c]{@{}c@{}}Performance \\ gain in \%\end{tabular}} & \multicolumn{1}{c|}{\begin{tabular}[c]{@{}c@{}}Additional\\ energy\\ usage in \%\end{tabular}} & \multicolumn{1}{c|}{\begin{tabular}[c]{@{}c@{}}Change in\\ MOPS / Watt\\ in \%\end{tabular}} & \multicolumn{1}{c|}{\begin{tabular}[c]{@{}c@{}}Performance\\ gain in \%\end{tabular}} & \multicolumn{1}{c|}{\begin{tabular}[c]{@{}c@{}}Additional\\ energy\\ usage in \%\end{tabular}} & \multicolumn{1}{c|}{\begin{tabular}[c]{@{}c@{}}Change in\\ MOPS / Watt\\ in \%\end{tabular}} & \multicolumn{1}{c|}{\begin{tabular}[c]{@{}c@{}}Performance\\ gain in \%\end{tabular}} & \multicolumn{1}{c|}{\begin{tabular}[c]{@{}c@{}}Additional\\ energy\\ usage in \%\end{tabular}} & \multicolumn{1}{c|}{\begin{tabular}[c]{@{}c@{}}Change in\\ MOPS / Watt\\ in \%\end{tabular}} \\ \hline
\multicolumn{1}{|l|}{CG-2-2S / CG-2-1S}                                                                              & \multicolumn{1}{r|}{9.64}                                                              & \multicolumn{1}{r|}{50.36}                                                                     & \multicolumn{1}{r|}{-33.49}                                                                   & \multicolumn{1}{r|}{8.92}                                                             & \multicolumn{1}{r|}{52.85}                                                                     & \multicolumn{1}{r|}{-34.57}                                                                   & \multicolumn{1}{r|}{13.50}                                                            & \multicolumn{1}{r|}{47.20}                                                                     & \multicolumn{1}{r|}{-32.06}                                                                   \\ \hline
\multicolumn{1}{|l|}{CG-4-2S / CG-4-1S}                                                                              & \multicolumn{1}{r|}{16.53}                                                             & \multicolumn{1}{r|}{42.22}                                                                     & \multicolumn{1}{r|}{-29.69}                                                                   & \multicolumn{1}{r|}{15.84}                                                            & \multicolumn{1}{r|}{44.42}                                                                     & \multicolumn{1}{r|}{-30.76}                                                                   & \multicolumn{1}{r|}{19.18}                                                            & \multicolumn{1}{r|}{40.03}                                                                     & \multicolumn{1}{r|}{-28.59}                                                                   \\ \hline
\multicolumn{1}{|l|}{CG-8-2S / CG-8-1S}                                                                              & \multicolumn{1}{r|}{15.71}                                                             & \multicolumn{1}{r|}{22.83}                                                                     & \multicolumn{1}{r|}{-18.58}                                                                   & \multicolumn{1}{r|}{12.07}                                                            & \multicolumn{1}{r|}{28.63}                                                                     & \multicolumn{1}{r|}{-22.26}                                                                   & \multicolumn{1}{r|}{16.14}                                                            & \multicolumn{1}{r|}{23.85}                                                                     & \multicolumn{1}{r|}{-19.24}                                                                   \\ \hline
                                                                                                                   & \multicolumn{1}{l}{}                                                                   & \multicolumn{1}{l}{}                                                                           & \multicolumn{1}{l}{}                                                                          & \multicolumn{1}{l}{}                                                                  & \multicolumn{1}{l}{}                                                                           & \multicolumn{1}{l}{}                                                                          & \multicolumn{1}{l}{}                                                                  & \multicolumn{1}{l}{}                                                                           & \multicolumn{1}{l}{}                                                                          \\ \cline{2-10} 
\multicolumn{1}{l|}{\textbf{\begin{tabular}[c]{@{}l@{}}Second socket\\ always included\\ even when not used\end{tabular}}}                                                                          & \multicolumn{3}{c|}{MPI}                                                                                                                                                                                                                                                                & \multicolumn{3}{c|}{OpenMP}                                                                                                                                                                                                                                                            & \multicolumn{3}{c|}{UPC}                                                                                                                                                                                                                                                               \\ \hline
\multicolumn{1}{|c|}{CG}                                                                                           & \multicolumn{1}{c|}{\begin{tabular}[c]{@{}c@{}}Performance\\ gain in \%\end{tabular}}  & \multicolumn{1}{c|}{\begin{tabular}[c]{@{}c@{}}Additional\\ energy\\ usage in \%\end{tabular}} & \multicolumn{1}{c|}{\begin{tabular}[c]{@{}c@{}}Change in\\ MOPS / Watt\\ in \%\end{tabular}} & \multicolumn{1}{c|}{\begin{tabular}[c]{@{}c@{}}Performance\\ gain in \%\end{tabular}} & \multicolumn{1}{c|}{\begin{tabular}[c]{@{}c@{}}Additional\\ energy\\ usage in \%\end{tabular}} & \multicolumn{1}{c|}{\begin{tabular}[c]{@{}c@{}}Change in\\ MOPS / Watt\\ in \%\end{tabular}} & \multicolumn{1}{c|}{\begin{tabular}[c]{@{}c@{}}Performance\\ gain in \%\end{tabular}} & \multicolumn{1}{c|}{\begin{tabular}[c]{@{}c@{}}Additional\\ energy\\ usage in \%\end{tabular}} & \multicolumn{1}{c|}{\begin{tabular}[c]{@{}c@{}}Change in\\ MOPS / Watt\\ in \%\end{tabular}} \\ \hline
\multicolumn{1}{|l|}{CG-2-2S / CG-2-1S}                                                                              & \multicolumn{1}{r|}{9.64}                                                              & \multicolumn{1}{r|}{8.86}                                                                      & \multicolumn{1}{r|}{-8.14}                                                                    & \multicolumn{1}{r|}{8.92}                                                             & \multicolumn{1}{r|}{10.02}                                                                     & \multicolumn{1}{r|}{-9.10}                                                                    & \multicolumn{1}{r|}{13.50}                                                            & \multicolumn{1}{r|}{5.28}                                                                      & \multicolumn{1}{r|}{-5.01}                                                                    \\ \hline
\multicolumn{1}{|l|}{CG-4-2S / CG-4-1S}                                                                              & \multicolumn{1}{r|}{16.53}                                                             & \multicolumn{1}{r|}{7.83}                                                                      & \multicolumn{1}{r|}{-7.26}                                                                    & \multicolumn{1}{r|}{15.84}                                                            & \multicolumn{1}{r|}{8.68}                                                                      & \multicolumn{1}{r|}{-8.00}                                                                    & \multicolumn{1}{r|}{19.18}                                                            & \multicolumn{1}{r|}{4.79}                                                                      & \multicolumn{1}{r|}{-4.58}                                                                    \\ \hline
\multicolumn{1}{|l|}{CG-8-2S / CG-8-1S}                                                                              & \multicolumn{1}{r|}{15.71}                                                             & \multicolumn{1}{r|}{0.63}                                                                      & \multicolumn{1}{r|}{-0.62}                                                                    & \multicolumn{1}{r|}{12.07}                                                            & \multicolumn{1}{r|}{3.55}                                                                      & \multicolumn{1}{r|}{-3.43}                                                                    & \multicolumn{1}{r|}{16.14}                                                            & \multicolumn{1}{r|}{0.23}                                                                      & \multicolumn{1}{r|}{-0.21}                                                                    \\ \hline
\end{tabular}
}
\end{table*}


Even though there is no global \textsl{winner} in the obtained single-node  measurements, UPC is able to compete with both OpenMP and MPI.
UPC arrives in second place for CG-8-1S, CG-8-2S, CG-16-2S, IS-2-2S, IS-4-2S, IS-8-2S and MG-4-1S.

As described in \cite{shan2010programming}, on a single-node platform, UPC scales well over more CPU cores and competes well with OpenMP and MPI. However, the performance of MPI or OpenMP is better in many cases.

For general-purpose architectures with high static power such as Intel Sandy Bridge, energy efficiency is directly connected to performance results. Therefore, the best results in energy efficiency are achieved, in most cases, for the kernels and thread-counts delivering the highest performance in MOPS.

However, by looking closely at the performance and performance per watt, it is possible to highlight that running a program in MPI, OpenMP or UPC over only 1 socket instead of two, when it is possible, delivers better energy efficiency. In Table \ref{table:energyTable}, we report values for CG in order to show a comparison between runs over 2,4 and 8 cores using 1 or 2 sockets. Table \ref{table:energyTable} is divided in two parts: the first part where the measurements on 1 socket do not include the measurement of the second idling socket, the second part where the idle socket measurements are included in the reported values. Table \ref{table:energyTable} corresponds to Figure \ref{fig:cgMops}, we chose to only represent the value of CG because the results of CG in terms of energy efficiency difference between one socket and two sockets are representative of all the kernels.
In the first part of Table \ref{table:energyTable}, we can see that CG-2-2S in MPI (CG running over two threads using two sockets) delivers better performance  than CG-2-1S in MPI (CG running over two threads using one socket): +9.64\% MOPS, but it costs +50,36\% in energy (Joules) and delivers -33.49\% MOPS per Watt.
In the second part of Table \ref{table:energyTable}, we can see that the same comparison, including values from the second idle socket, gives a similar conclusion as before: using two sockets consumes +8.86\% energy and delivers -8.14\% in MOPS per Watt.
These observations are also visible in Figure \ref{fig:cgMops}, \ref{fig:mgMops}, \ref{fig:ftMops} and \ref{fig:isMops}: the lines representing the MOPS per Watt are higher for kernels using one socket for equivalent threads/process count than that of kernels using two sockets.

\section{Discussion}
\label{discussion}
\begin{table*}[t]
\centering
\caption{Memory Traffic, L3 cache hit ration and L2 cache hit ratio, for CG, MG, FT and IS implemented in MPI, OpenMP and UPC \newline Results are presented for 2,4 and 8 cores using one and two sockets and 16 cores using two sockets}
\label{perfDetail}
\scalebox{0.98}{
\begin{tabular}{lrrrrrrrrrrrr}
\cline{2-13}
\multicolumn{1}{l|}{}                        & \multicolumn{4}{c|}{MPI}                                                                                                                                                                                                                                                                                                                           & \multicolumn{4}{c|}{OpenMP}                                                                                                                                                                                                                                                                                                                        & \multicolumn{4}{c|}{UPC}                                                                                                                                                                                                                                                                                                                           \\ \hline
\multicolumn{1}{|c|}{CG - Size C}            & \multicolumn{1}{c|}{\begin{tabular}[c]{@{}c@{}}Mem\\ traffic\\ (GB)\end{tabular}} & \multicolumn{1}{c|}{\begin{tabular}[c]{@{}c@{}}L3 hit\\ ratio\end{tabular}} & \multicolumn{1}{c|}{\begin{tabular}[c]{@{}c@{}}L3\\ access\\ $\times 10^6$\end{tabular}} & \multicolumn{1}{c|}{\begin{tabular}[c]{@{}c@{}}L2 hit\\ ratio\end{tabular}} & \multicolumn{1}{c|}{\begin{tabular}[c]{@{}c@{}}Mem\\ traffic\\ (GB)\end{tabular}} & \multicolumn{1}{c|}{\begin{tabular}[c]{@{}c@{}}L3 hit\\ ratio\end{tabular}} & \multicolumn{1}{c|}{\begin{tabular}[c]{@{}c@{}}L3\\ access\\ $\times 10^6$\end{tabular}} & \multicolumn{1}{c|}{\begin{tabular}[c]{@{}c@{}}L2 hit\\ ratio\end{tabular}} & \multicolumn{1}{c|}{\begin{tabular}[c]{@{}c@{}}Mem\\ traffic\\ (GB)\end{tabular}} & \multicolumn{1}{c|}{\begin{tabular}[c]{@{}c@{}}L3 hit\\ ratio\end{tabular}} & \multicolumn{1}{c|}{\begin{tabular}[c]{@{}c@{}}L3\\ access\\ $\times 10^6$\end{tabular}} & \multicolumn{1}{c|}{\begin{tabular}[c]{@{}c@{}}L2 hit\\ ratio\end{tabular}} \\ \hline
\multicolumn{1}{|l|}{1 Threads - 1 socket}   & \multicolumn{1}{r|}{833.52}                                                       & \multicolumn{1}{r|}{0.81}                                                   & \multicolumn{1}{r|}{16049}                                                                         & \multicolumn{1}{r|}{0.14}                                                   & \multicolumn{1}{r|}{834.71}                                                       & \multicolumn{1}{r|}{0.81}                                                   & \multicolumn{1}{r|}{16049}                                                                         & \multicolumn{1}{r|}{0.14}                                                   & \multicolumn{1}{r|}{834.58}                                                       & \multicolumn{1}{r|}{0.81}                                                   & \multicolumn{1}{r|}{16049}                                                                         & \multicolumn{1}{r|}{0.14}                                                   \\ \hline
\multicolumn{1}{|l|}{2 Threads - 1 socket}   & \multicolumn{1}{r|}{846.14}                                                       & \multicolumn{1}{r|}{0.77}                                                   & \multicolumn{1}{r|}{16883}                                                                         & \multicolumn{1}{r|}{0.27}                                                   & \multicolumn{1}{r|}{836.03}                                                       & \multicolumn{1}{r|}{0.81}                                                   & \multicolumn{1}{r|}{16049}                                                                         & \multicolumn{1}{r|}{0.14}                                                   & \multicolumn{1}{r|}{856.01}                                                       & \multicolumn{1}{r|}{0.77}                                                   & \multicolumn{1}{r|}{16883}                                                                         & \multicolumn{1}{r|}{0.27}                                                   \\ \hline
\multicolumn{1}{|l|}{2 Threads - 2 sockets}  & \multicolumn{1}{r|}{845.22}                                                       & \multicolumn{1}{r|}{0.77}                                                   & \multicolumn{1}{r|}{17391}                                                                         & \multicolumn{1}{r|}{0.27}                                                   & \multicolumn{1}{r|}{841.93}                                                       & \multicolumn{1}{r|}{0.81}                                                   & \multicolumn{1}{r|}{16798}                                                                         & \multicolumn{1}{r|}{0.14}                                                   & \multicolumn{1}{r|}{854.4}                                                        & \multicolumn{1}{r|}{0.77}                                                   & \multicolumn{1}{r|}{17725}                                                                         & \multicolumn{1}{r|}{0.27}                                                   \\ \hline
\multicolumn{1}{|l|}{4 Threads - 1 socket}   & \multicolumn{1}{r|}{854.25}                                                       & \multicolumn{1}{r|}{0.77}                                                   & \multicolumn{1}{r|}{16883}                                                                         & \multicolumn{1}{r|}{0.27}                                                   & \multicolumn{1}{r|}{834.26}                                                       & \multicolumn{1}{r|}{0.81}                                                   & \multicolumn{1}{r|}{16049}                                                                         & \multicolumn{1}{r|}{0.14}                                                   & \multicolumn{1}{r|}{878.64}                                                       & \multicolumn{1}{r|}{0.77}                                                   & \multicolumn{1}{r|}{16883}                                                                         & \multicolumn{1}{r|}{0.27}                                                   \\ \hline
\multicolumn{1}{|l|}{4 Threads - 2 sockets}  & \multicolumn{1}{r|}{859.01}                                                       & \multicolumn{1}{r|}{0.77}                                                   & \multicolumn{1}{r|}{17474}                                                                         & \multicolumn{1}{r|}{0.27}                                                   & \multicolumn{1}{r|}{842.96}                                                       & \multicolumn{1}{r|}{0.81}                                                   & \multicolumn{1}{r|}{16823}                                                                         & \multicolumn{1}{r|}{0.14}                                                   & \multicolumn{1}{r|}{881.10}                                                       & \multicolumn{1}{r|}{0.77}                                                   & \multicolumn{1}{r|}{17817}                                                                         & \multicolumn{1}{r|}{0.27}                                                   \\ \hline
\multicolumn{1}{|l|}{8 Threads - 1 socket}   & \multicolumn{1}{r|}{878.89}                                                       & \multicolumn{1}{r|}{0.74}                                                   & \multicolumn{1}{r|}{11345}                                                                         & \multicolumn{1}{r|}{0.56}                                                   & \multicolumn{1}{r|}{832.17}                                                       & \multicolumn{1}{r|}{0.81}                                                   & \multicolumn{1}{r|}{16049}                                                                         & \multicolumn{1}{r|}{0.14}                                                   & \multicolumn{1}{r|}{909.09}                                                       & \multicolumn{1}{r|}{0.73}                                                   & \multicolumn{1}{r|}{12042}                                                                         & \multicolumn{1}{r|}{0.56}                                                   \\ \hline
\multicolumn{1}{|l|}{8 Threads - 2 sockets}  & \multicolumn{1}{r|}{881.81}                                                       & \multicolumn{1}{r|}{0.74}                                                   & \multicolumn{1}{r|}{11211}                                                                         & \multicolumn{1}{r|}{0.57}                                                   & \multicolumn{1}{r|}{841.31}                                                       & \multicolumn{1}{r|}{0.80}                                                   & \multicolumn{1}{r|}{17023}                                                                         & \multicolumn{1}{r|}{0.14}                                                   & \multicolumn{1}{r|}{912.42}                                                       & \multicolumn{1}{r|}{0.73}                                                   & \multicolumn{1}{r|}{11827}                                                                         & \multicolumn{1}{r|}{0.56}                                                   \\ \hline
\multicolumn{1}{|l|}{16 Threads - 2 sockets} & \multicolumn{1}{r|}{908.34}                                                       & \multicolumn{1}{r|}{0.74}                                                   & \multicolumn{1}{r|}{11535}                                                                         & \multicolumn{1}{r|}{0.565}                                                  & \multicolumn{1}{r|}{840.23}                                                       & \multicolumn{1}{r|}{0.80}                                                   & \multicolumn{1}{r|}{17014}                                                                         & \multicolumn{1}{r|}{0.14}                                                   & \multicolumn{1}{r|}{968.8}                                                        & \multicolumn{1}{r|}{0.72}                                                   & \multicolumn{1}{r|}{12439}                                                                         & \multicolumn{1}{r|}{0.56}                                                   \\ \hline
                                             & \multicolumn{1}{l}{}                                                              & \multicolumn{1}{l}{}                                                        & \multicolumn{1}{l}{}                                                                               & \multicolumn{1}{l}{}                                                        & \multicolumn{1}{l}{}                                                              & \multicolumn{1}{l}{}                                                        & \multicolumn{1}{l}{}                                                                               & \multicolumn{1}{l}{}                                                        & \multicolumn{1}{l}{}                                                              & \multicolumn{1}{l}{}                                                        & \multicolumn{1}{l}{}                                                                               & \multicolumn{1}{l}{}                                                        \\
                                             & \multicolumn{1}{l}{}                                                              & \multicolumn{1}{l}{}                                                        & \multicolumn{1}{l}{}                                                                               & \multicolumn{1}{l}{}                                                        & \multicolumn{1}{l}{}                                                              & \multicolumn{1}{l}{}                                                        & \multicolumn{1}{l}{}                                                                               & \multicolumn{1}{l}{}                                                        & \multicolumn{1}{l}{}                                                              & \multicolumn{1}{l}{}                                                        & \multicolumn{1}{l}{}                                                                               & \multicolumn{1}{l}{}                                                        \\ \cline{2-13} 
\multicolumn{1}{l|}{}                        & \multicolumn{4}{c|}{MPI}                                                                                                                                                                                                                                                                                                                           & \multicolumn{4}{c|}{OpenMP}                                                                                                                                                                                                                                                                                                                        & \multicolumn{4}{c|}{UPC}                                                                                                                                                                                                                                                                                                                           \\ \hline
\multicolumn{1}{|c|}{MG - Size C}            & \multicolumn{1}{c|}{\begin{tabular}[c]{@{}c@{}}Mem\\ traffic\\ (GB)\end{tabular}} & \multicolumn{1}{c|}{\begin{tabular}[c]{@{}c@{}}L3 hit\\ ratio\end{tabular}} & \multicolumn{1}{c|}{\begin{tabular}[c]{@{}c@{}}L3\\ access\\ $\times 10^6$\end{tabular}} & \multicolumn{1}{c|}{\begin{tabular}[c]{@{}c@{}}L2 hit\\ ratio\end{tabular}} & \multicolumn{1}{c|}{\begin{tabular}[c]{@{}c@{}}Mem\\ traffic\\ (GB)\end{tabular}} & \multicolumn{1}{c|}{\begin{tabular}[c]{@{}c@{}}L3 hit\\ ratio\end{tabular}} & \multicolumn{1}{c|}{\begin{tabular}[c]{@{}c@{}}L3\\ access\\ $\times 10^6$\end{tabular}} & \multicolumn{1}{c|}{\begin{tabular}[c]{@{}c@{}}L2 hit\\ ratio\end{tabular}} & \multicolumn{1}{c|}{\begin{tabular}[c]{@{}c@{}}Mem\\ traffic\\ (GB)\end{tabular}} & \multicolumn{1}{c|}{\begin{tabular}[c]{@{}c@{}}L3 hit\\ ratio\end{tabular}} & \multicolumn{1}{c|}{\begin{tabular}[c]{@{}c@{}}L3\\ access\\ $\times 10^6$\end{tabular}} & \multicolumn{1}{c|}{\begin{tabular}[c]{@{}c@{}}L2 hit\\ ratio\end{tabular}} \\ \hline
\multicolumn{1}{|l|}{1 Threads - 1 socket}   & \multicolumn{1}{r|}{388.53}                                                       & \multicolumn{1}{r|}{0.38}                                                   & \multicolumn{1}{r|}{1174}                                                                          & \multicolumn{1}{r|}{0.83}                                                   & \multicolumn{1}{r|}{372.25}                                                       & \multicolumn{1}{r|}{0.31}                                                   & \multicolumn{1}{r|}{1103}                                                                          & \multicolumn{1}{r|}{0.87}                                                   & \multicolumn{1}{r|}{369.09}                                                       & \multicolumn{1}{r|}{0.52}                                                   & \multicolumn{1}{r|}{810}                                                                           & \multicolumn{1}{r|}{0.78}                                                   \\ \hline
\multicolumn{1}{|l|}{2 Threads - 1 socket}   & \multicolumn{1}{r|}{459.01}                                                       & \multicolumn{1}{r|}{0.35}                                                   & \multicolumn{1}{r|}{1540}                                                                          & \multicolumn{1}{r|}{0.82}                                                   & \multicolumn{1}{r|}{437.43}                                                       & \multicolumn{1}{r|}{0.28}                                                   & \multicolumn{1}{r|}{1439}                                                                          & \multicolumn{1}{r|}{0.85}                                                   & \multicolumn{1}{r|}{435.71}                                                       & \multicolumn{1}{r|}{0.48}                                                   & \multicolumn{1}{r|}{1100}                                                                          & \multicolumn{1}{r|}{0.75}                                                   \\ \hline
\multicolumn{1}{|l|}{2 Threads - 2 sockets}  & \multicolumn{1}{r|}{383.96}                                                       & \multicolumn{1}{r|}{0.42}                                                   & \multicolumn{1}{r|}{1005}                                                                          & \multicolumn{1}{r|}{0.83}                                                   & \multicolumn{1}{r|}{370.92}                                                       & \multicolumn{1}{r|}{0.31}                                                   & \multicolumn{1}{r|}{1084}                                                                          & \multicolumn{1}{r|}{0.87}                                                   & \multicolumn{1}{r|}{370.87}                                                       & \multicolumn{1}{r|}{0.53}                                                   & \multicolumn{1}{r|}{794}                                                                           & \multicolumn{1}{r|}{0.78}                                                   \\ \hline
\multicolumn{1}{|l|}{4 Threads - 1 socket}   & \multicolumn{1}{r|}{545.86}                                                       & \multicolumn{1}{r|}{0.26}                                                   & \multicolumn{1}{r|}{3031}                                                                          & \multicolumn{1}{r|}{0.76}                                                   & \multicolumn{1}{r|}{513.15}                                                       & \multicolumn{1}{r|}{0.23}                                                   & \multicolumn{1}{r|}{2078}                                                                          & \multicolumn{1}{r|}{0.84}                                                   & \multicolumn{1}{r|}{436.42}                                                       & \multicolumn{1}{r|}{0.46}                                                   & \multicolumn{1}{r|}{1178}                                                                          & \multicolumn{1}{r|}{0.75}                                                   \\ \hline
\multicolumn{1}{|l|}{4 Threads - 2 sockets}  & \multicolumn{1}{r|}{406.04}                                                       & \multicolumn{1}{r|}{0.39}                                                   & \multicolumn{1}{r|}{1275}                                                                          & \multicolumn{1}{r|}{0.81}                                                   & \multicolumn{1}{r|}{434.59}                                                       & \multicolumn{1}{r|}{0.27}                                                   & \multicolumn{1}{r|}{1448}                                                                          & \multicolumn{1}{r|}{0.86}                                                   & \multicolumn{1}{r|}{373.88}                                                       & \multicolumn{1}{r|}{0.52}                                                   & \multicolumn{1}{r|}{850}                                                                           & \multicolumn{1}{r|}{0.77}                                                   \\ \hline
\multicolumn{1}{|l|}{8 Threads - 1 socket}   & \multicolumn{1}{r|}{476.24}                                                       & \multicolumn{1}{r|}{0.20}                                                   & \multicolumn{1}{r|}{4355}                                                                          & \multicolumn{1}{r|}{0.33}                                                   & \multicolumn{1}{r|}{521.27}                                                       & \multicolumn{1}{r|}{0.20}                                                   & \multicolumn{1}{r|}{3415}                                                                          & \multicolumn{1}{r|}{0.79}                                                   & \multicolumn{1}{r|}{453.60}                                                       & \multicolumn{1}{r|}{0.28}                                                   & \multicolumn{1}{r|}{2843}                                                                          & \multicolumn{1}{r|}{0.49}                                                   \\ \hline
\multicolumn{1}{|l|}{8 Threads - 2 sockets}  & \multicolumn{1}{r|}{400.37}                                                       & \multicolumn{1}{r|}{0.23}                                                   & \multicolumn{1}{r|}{2722}                                                                          & \multicolumn{1}{r|}{0.38}                                                   & \multicolumn{1}{r|}{510.78}                                                       & \multicolumn{1}{r|}{0.23}                                                   & \multicolumn{1}{r|}{2061}                                                                          & \multicolumn{1}{r|}{0.84}                                                   & \multicolumn{1}{r|}{388.32}                                                       & \multicolumn{1}{r|}{0.33}                                                   & \multicolumn{1}{r|}{1861}                                                                          & \multicolumn{1}{r|}{0.57}                                                   \\ \hline
\multicolumn{1}{|l|}{16 Threads - 2 sockets} & \multicolumn{1}{r|}{487.40}                                                       & \multicolumn{1}{r|}{0.20}                                                   & \multicolumn{1}{r|}{4390}                                                                          & \multicolumn{1}{r|}{0.32}                                                   & \multicolumn{1}{r|}{520.69}                                                       & \multicolumn{1}{r|}{0.19}                                                   & \multicolumn{1}{r|}{3663}                                                                          & \multicolumn{1}{r|}{0.795}                                                  & \multicolumn{1}{r|}{452.36}                                                       & \multicolumn{1}{r|}{0.28}                                                   & \multicolumn{1}{r|}{2782}                                                                          & \multicolumn{1}{r|}{0.49}                                                   \\ \hline
                                             & \multicolumn{1}{l}{}                                                              & \multicolumn{1}{l}{}                                                        & \multicolumn{1}{l}{}                                                                               & \multicolumn{1}{l}{}                                                        & \multicolumn{1}{l}{}                                                              & \multicolumn{1}{l}{}                                                        & \multicolumn{1}{l}{}                                                                               & \multicolumn{1}{l}{}                                                        & \multicolumn{1}{l}{}                                                              & \multicolumn{1}{l}{}                                                        & \multicolumn{1}{l}{}                                                                               & \multicolumn{1}{l}{}                                                        \\
                                             & \multicolumn{1}{l}{}                                                              & \multicolumn{1}{l}{}                                                        & \multicolumn{1}{l}{}                                                                               & \multicolumn{1}{l}{}                                                        & \multicolumn{1}{l}{}                                                              & \multicolumn{1}{l}{}                                                        & \multicolumn{1}{l}{}                                                                               & \multicolumn{1}{l}{}                                                        & \multicolumn{1}{l}{}                                                              & \multicolumn{1}{l}{}                                                        & \multicolumn{1}{l}{}                                                                               & \multicolumn{1}{l}{}                                                        \\ \cline{2-13} 
\multicolumn{1}{l|}{}                        & \multicolumn{4}{c|}{MPI}                                                                                                                                                                                                                                                                                                                           & \multicolumn{4}{c|}{OpenMP}                                                                                                                                                                                                                                                                                                                        & \multicolumn{4}{c|}{UPC}                                                                                                                                                                                                                                                                                                                           \\ \hline
\multicolumn{1}{|c|}{FT - Size C}            & \multicolumn{1}{c|}{\begin{tabular}[c]{@{}c@{}}Mem\\ traffic\\ (GB)\end{tabular}} & \multicolumn{1}{c|}{\begin{tabular}[c]{@{}c@{}}L3 hit\\ ratio\end{tabular}} & \multicolumn{1}{c|}{\begin{tabular}[c]{@{}c@{}}L3\\ access\\ $\times 10^6$\end{tabular}} & \multicolumn{1}{c|}{\begin{tabular}[c]{@{}c@{}}L2 hit\\ ratio\end{tabular}} & \multicolumn{1}{c|}{\begin{tabular}[c]{@{}c@{}}Mem\\ traffic\\ (GB)\end{tabular}} & \multicolumn{1}{c|}{\begin{tabular}[c]{@{}c@{}}L3 hit\\ ratio\end{tabular}} & \multicolumn{1}{c|}{\begin{tabular}[c]{@{}c@{}}L3\\ access\\ $\times 10^6$\end{tabular}} & \multicolumn{1}{c|}{\begin{tabular}[c]{@{}c@{}}L2 hit\\ ratio\end{tabular}} & \multicolumn{1}{c|}{\begin{tabular}[c]{@{}c@{}}Mem\\ traffic\\ (GB)\end{tabular}} & \multicolumn{1}{c|}{\begin{tabular}[c]{@{}c@{}}L3 hit\\ ratio\end{tabular}} & \multicolumn{1}{c|}{\begin{tabular}[c]{@{}c@{}}L3\\ access\\ $\times 10^6$\end{tabular}} & \multicolumn{1}{c|}{\begin{tabular}[c]{@{}c@{}}L2 hit\\ ratio\end{tabular}} \\ \hline
\multicolumn{1}{|l|}{1 Threads - 1 socket}   & \multicolumn{1}{r|}{617.90}                                                       & \multicolumn{1}{r|}{0.38}                                                   & \multicolumn{1}{r|}{8905}                                                                          & \multicolumn{1}{r|}{0.56}                                                   & \multicolumn{1}{r|}{471.60}                                                       & \multicolumn{1}{r|}{0.79}                                                   & \multicolumn{1}{r|}{1400}                                                                          & \multicolumn{1}{r|}{0.43}                                                   & \multicolumn{1}{r|}{}                                                             & \multicolumn{1}{r|}{}                                                       & \multicolumn{1}{r|}{}                                                                              & \multicolumn{1}{r|}{}                                                       \\ \hline
\multicolumn{1}{|l|}{2 Threads - 1 socket}   & \multicolumn{1}{r|}{852.26}                                                       & \multicolumn{1}{r|}{0.50}                                                   & \multicolumn{1}{r|}{6244}                                                                          & \multicolumn{1}{r|}{0.63}                                                   & \multicolumn{1}{r|}{470.57}                                                       & \multicolumn{1}{r|}{0.79}                                                   & \multicolumn{1}{r|}{1437}                                                                          & \multicolumn{1}{r|}{0.42}                                                   & \multicolumn{1}{r|}{958.65}                                                       & \multicolumn{1}{r|}{0.31}                                                   & \multicolumn{1}{r|}{16610}                                                                         & \multicolumn{1}{r|}{0.56}                                                   \\ \hline
\multicolumn{1}{|l|}{2 Threads - 2 sockets}  & \multicolumn{1}{r|}{853.49}                                                       & \multicolumn{1}{r|}{0.51}                                                   & \multicolumn{1}{r|}{6063}                                                                          & \multicolumn{1}{r|}{0.63}                                                   & \multicolumn{1}{r|}{471.54}                                                       & \multicolumn{1}{r|}{0.80}                                                   & \multicolumn{1}{r|}{1378}                                                                          & \multicolumn{1}{r|}{0.41}                                                   & \multicolumn{1}{r|}{959.68}                                                       & \multicolumn{1}{r|}{0.31}                                                   & \multicolumn{1}{r|}{16587}                                                                         & \multicolumn{1}{r|}{0.57}                                                   \\ \hline
\multicolumn{1}{|l|}{4 Threads - 1 socket}   & \multicolumn{1}{r|}{867.25}                                                       & \multicolumn{1}{r|}{0.49}                                                   & \multicolumn{1}{r|}{6753}                                                                          & \multicolumn{1}{r|}{0.62}                                                   & \multicolumn{1}{r|}{475.96}                                                       & \multicolumn{1}{r|}{0.79}                                                   & \multicolumn{1}{r|}{1446}                                                                          & \multicolumn{1}{r|}{0.41}                                                   & \multicolumn{1}{r|}{960.74}                                                       & \multicolumn{1}{r|}{0.31}                                                   & \multicolumn{1}{r|}{16597}                                                                         & \multicolumn{1}{r|}{0.59}                                                   \\ \hline
\multicolumn{1}{|l|}{4 Threads - 2 sockets}  & \multicolumn{1}{r|}{865.76}                                                       & \multicolumn{1}{r|}{0.50}                                                   & \multicolumn{1}{r|}{6291}                                                                          & \multicolumn{1}{r|}{0.64}                                                   & \multicolumn{1}{r|}{470.93}                                                       & \multicolumn{1}{r|}{0.79}                                                   & \multicolumn{1}{r|}{1424}                                                                          & \multicolumn{1}{r|}{0.41}                                                   & \multicolumn{1}{r|}{960.21}                                                       & \multicolumn{1}{r|}{0.31}                                                   & \multicolumn{1}{r|}{16432}                                                                         & \multicolumn{1}{r|}{0.58}                                                   \\ \hline
\multicolumn{1}{|l|}{8 Threads - 1 socket}   & \multicolumn{1}{r|}{881.36}                                                       & \multicolumn{1}{r|}{0.42}                                                   & \multicolumn{1}{r|}{8767}                                                                          & \multicolumn{1}{r|}{0.64}                                                   & \multicolumn{1}{r|}{491.11}                                                       & \multicolumn{1}{r|}{0.76}                                                   & \multicolumn{1}{r|}{1601}                                                                          & \multicolumn{1}{r|}{0.43}                                                   & \multicolumn{1}{r|}{977.94}                                                       & \multicolumn{1}{r|}{0.28}                                                   & \multicolumn{1}{r|}{19582}                                                                         & \multicolumn{1}{r|}{0.56}                                                   \\ \hline
\multicolumn{1}{|l|}{8 Threads - 2 sockets}  & \multicolumn{1}{r|}{873.64}                                                       & \multicolumn{1}{r|}{0.48}                                                   & \multicolumn{1}{r|}{6941}                                                                          & \multicolumn{1}{r|}{0.64}                                                   & \multicolumn{1}{r|}{475.76}                                                       & \multicolumn{1}{r|}{0.78}                                                   & \multicolumn{1}{r|}{1463}                                                                          & \multicolumn{1}{r|}{0.42}                                                   & \multicolumn{1}{r|}{975.48}                                                       & \multicolumn{1}{r|}{0.30}                                                   & \multicolumn{1}{r|}{17667}                                                                         & \multicolumn{1}{r|}{0.57}                                                   \\ \hline
\multicolumn{1}{|l|}{16 Threads - 2 sockets} & \multicolumn{1}{r|}{878.80}                                                       & \multicolumn{1}{r|}{0.43}                                                   & \multicolumn{1}{r|}{8319}                                                                          & \multicolumn{1}{r|}{0.64}                                                   & \multicolumn{1}{r|}{497.33}                                                       & \multicolumn{1}{r|}{0.76}                                                   & \multicolumn{1}{r|}{1681}                                                                          & \multicolumn{1}{r|}{0.42}                                                   & \multicolumn{1}{r|}{1009}                                                         & \multicolumn{1}{r|}{0.26}                                                   & \multicolumn{1}{r|}{22562}                                                                         & \multicolumn{1}{r|}{0.56}                                                   \\ \hline
                                             & \multicolumn{1}{l}{}                                                              & \multicolumn{1}{l}{}                                                        & \multicolumn{1}{l}{}                                                                               & \multicolumn{1}{l}{}                                                        & \multicolumn{1}{l}{}                                                              & \multicolumn{1}{l}{}                                                        & \multicolumn{1}{l}{}                                                                               & \multicolumn{1}{l}{}                                                        & \multicolumn{1}{l}{}                                                              & \multicolumn{1}{l}{}                                                        & \multicolumn{1}{l}{}                                                                               & \multicolumn{1}{l}{}                                                        \\
                                             & \multicolumn{1}{l}{}                                                              & \multicolumn{1}{l}{}                                                        & \multicolumn{1}{l}{}                                                                               & \multicolumn{1}{l}{}                                                        & \multicolumn{1}{l}{}                                                              & \multicolumn{1}{l}{}                                                        & \multicolumn{1}{l}{}                                                                               & \multicolumn{1}{l}{}                                                        & \multicolumn{1}{l}{}                                                              & \multicolumn{1}{l}{}                                                        & \multicolumn{1}{l}{}                                                                               & \multicolumn{1}{l}{}                                                        \\ \cline{2-13} 
\multicolumn{1}{l|}{}                        & \multicolumn{4}{c|}{MPI}                                                                                                                                                                                                                                                                                                                           & \multicolumn{4}{c|}{OpenMP}                                                                                                                                                                                                                                                                                                                        & \multicolumn{4}{c|}{UPC}                                                                                                                                                                                                                                                                                                                           \\ \hline
\multicolumn{1}{|c|}{IS - Size C}            & \multicolumn{1}{c|}{\begin{tabular}[c]{@{}c@{}}Mem\\ traffic\\ (GB)\end{tabular}} & \multicolumn{1}{c|}{\begin{tabular}[c]{@{}c@{}}L3 hit\\ ratio\end{tabular}} & \multicolumn{1}{c|}{\begin{tabular}[c]{@{}c@{}}L3\\ access\\ $\times 10^6$\end{tabular}} & \multicolumn{1}{c|}{\begin{tabular}[c]{@{}c@{}}L2 hit\\ ratio\end{tabular}} & \multicolumn{1}{c|}{\begin{tabular}[c]{@{}c@{}}Mem\\ traffic\\ (GB)\end{tabular}} & \multicolumn{1}{c|}{\begin{tabular}[c]{@{}c@{}}L3 hit\\ ratio\end{tabular}} & \multicolumn{1}{c|}{\begin{tabular}[c]{@{}c@{}}L3\\ access\\ $\times 10^6$\end{tabular}} & \multicolumn{1}{c|}{\begin{tabular}[c]{@{}c@{}}L2 hit\\ ratio\end{tabular}} & \multicolumn{1}{c|}{\begin{tabular}[c]{@{}c@{}}Mem\\ traffic\\ (GB)\end{tabular}} & \multicolumn{1}{c|}{\begin{tabular}[c]{@{}c@{}}L3 hit\\ ratio\end{tabular}} & \multicolumn{1}{c|}{\begin{tabular}[c]{@{}c@{}}L3\\ access\\ $\times 10^6$\end{tabular}} & \multicolumn{1}{c|}{\begin{tabular}[c]{@{}c@{}}L2 hit\\ ratio\end{tabular}} \\ \hline
\multicolumn{1}{|l|}{1 Threads - 1 socket}   & \multicolumn{1}{r|}{}                                                             & \multicolumn{1}{r|}{}                                                       & \multicolumn{1}{r|}{}                                                                              & \multicolumn{1}{r|}{}                                                       & \multicolumn{1}{r|}{32.39}                                                        & \multicolumn{1}{r|}{0.03}                                                   & \multicolumn{1}{r|}{5600}                                                                          & \multicolumn{1}{r|}{0.63}                                                   & \multicolumn{1}{r|}{44.10}                                                        & \multicolumn{1}{r|}{0.07}                                                   & \multicolumn{1}{r|}{2629}                                                                          & \multicolumn{1}{r|}{0.62}                                                   \\ \hline
\multicolumn{1}{|l|}{2 Threads - 1 socket}   & \multicolumn{1}{r|}{47.67}                                                        & \multicolumn{1}{r|}{0.13}                                                   & \multicolumn{1}{r|}{1846}                                                                          & \multicolumn{1}{r|}{0.53}                                                   & \multicolumn{1}{r|}{32.23}                                                        & \multicolumn{1}{r|}{0.06}                                                   & \multicolumn{1}{r|}{2817}                                                                          & \multicolumn{1}{r|}{0.62}                                                   & \multicolumn{1}{r|}{45.68}                                                        & \multicolumn{1}{r|}{0.08}                                                   & \multicolumn{1}{r|}{2463}                                                                          & \multicolumn{1}{r|}{0.60}                                                   \\ \hline
\multicolumn{1}{|l|}{2 Threads - 2 sockets}  & \multicolumn{1}{r|}{47.52}                                                        & \multicolumn{1}{r|}{0.13}                                                   & \multicolumn{1}{r|}{1800}                                                                          & \multicolumn{1}{r|}{0.53}                                                   & \multicolumn{1}{r|}{33.84}                                                        & \multicolumn{1}{r|}{0.03}                                                   & \multicolumn{1}{r|}{5833}                                                                          & \multicolumn{1}{r|}{0.62}                                                   & \multicolumn{1}{r|}{45.75}                                                        & \multicolumn{1}{r|}{0.07}                                                   & \multicolumn{1}{r|}{2771}                                                                          & \multicolumn{1}{r|}{0.61}                                                   \\ \hline
\multicolumn{1}{|l|}{4 Threads - 1 socket}   & \multicolumn{1}{r|}{50.99}                                                        & \multicolumn{1}{r|}{0.13}                                                   & \multicolumn{1}{r|}{2115}                                                                          & \multicolumn{1}{r|}{0.49}                                                   & \multicolumn{1}{r|}{32.10}                                                        & \multicolumn{1}{r|}{0.06}                                                   & \multicolumn{1}{r|}{2833}                                                                          & \multicolumn{1}{r|}{0.62}                                                   & \multicolumn{1}{r|}{49.57}                                                        & \multicolumn{1}{r|}{0.06}                                                   & \multicolumn{1}{r|}{3900}                                                                          & \multicolumn{1}{r|}{0.56}                                                   \\ \hline
\multicolumn{1}{|l|}{4 Threads - 2 sockets}  & \multicolumn{1}{r|}{50.80}                                                        & \multicolumn{1}{r|}{0.13}                                                   & \multicolumn{1}{r|}{2092}                                                                          & \multicolumn{1}{r|}{0.50}                                                   & \multicolumn{1}{r|}{33.33}                                                        & \multicolumn{1}{r|}{0.05}                                                   & \multicolumn{1}{r|}{3867}                                                                          & \multicolumn{1}{r|}{0.62}                                                   & \multicolumn{1}{r|}{49.50}                                                        & \multicolumn{1}{r|}{0.07}                                                   & \multicolumn{1}{r|}{3523}                                                                          & \multicolumn{1}{r|}{0.57}                                                   \\ \hline
\multicolumn{1}{|l|}{8 Threads - 1 socket}   & \multicolumn{1}{r|}{56.27}                                                        & \multicolumn{1}{r|}{0.10}                                                   & \multicolumn{1}{r|}{3280}                                                                          & \multicolumn{1}{r|}{0.46}                                                   & \multicolumn{1}{r|}{32.22}                                                        & \multicolumn{1}{r|}{0.05}                                                   & \multicolumn{1}{r|}{3460}                                                                          & \multicolumn{1}{r|}{0.62}                                                   & \multicolumn{1}{r|}{57.78}                                                        & \multicolumn{1}{r|}{0.04}                                                   & \multicolumn{1}{r|}{8400}                                                                          & \multicolumn{1}{r|}{0.47}                                                   \\ \hline
\multicolumn{1}{|l|}{8 Threads - 2 sockets}  & \multicolumn{1}{r|}{55.68}                                                        & \multicolumn{1}{r|}{0.13}                                                   & \multicolumn{1}{r|}{2423}                                                                          & \multicolumn{1}{r|}{0.46}                                                   & \multicolumn{1}{r|}{33.08}                                                        & \multicolumn{1}{r|}{0.05}                                                   & \multicolumn{1}{r|}{3911}                                                                          & \multicolumn{1}{r|}{0.62}                                                   & \multicolumn{1}{r|}{56.84}                                                        & \multicolumn{1}{r|}{0.05}                                                   & \multicolumn{1}{r|}{6889}                                                                          & \multicolumn{1}{r|}{0.49}                                                   \\ \hline
\multicolumn{1}{|l|}{16 Threads - 2 sockets} & \multicolumn{1}{r|}{60.77}                                                        & \multicolumn{1}{r|}{0.13}                                                   & \multicolumn{1}{r|}{2815}                                                                          & \multicolumn{1}{r|}{0.42}                                                   & \multicolumn{1}{r|}{33.24}                                                        & \multicolumn{1}{r|}{0.04}                                                   & \multicolumn{1}{r|}{4550}                                                                          & \multicolumn{1}{r|}{0.61}                                                   & \multicolumn{1}{r|}{62.93}                                                        & \multicolumn{1}{r|}{0.07}                                                   & \multicolumn{1}{r|}{5143}                                                                          & \multicolumn{1}{r|}{0.44}                                                   \\ \hline
\end{tabular}
}
\end{table*}
In this section, we give an analysis of the differences in performance and energy efficiency that were observed in the previous section.

Intel PCM provides access to metrics such as memory traffic and hit and miss rates for L2 and L3 cache. In this section we will use measurements of these metrics to analyze the differences in performance and energy efficiency  among OpenMP, MPI, and UPC. Table \ref{perfDetail} shows the measurements obtained via Intel PCM. In Table \ref{perfDetail}, we use different metrics: Memory traffic ($read + write$) expressed in GigaBytes, L2 and L3 cache hit ratio, and L3 access given in millions of accesses. The results are given for each kernel (CG, MG, FT and IS) over 1, 2, 4 and 8 cores using one socket and 2, 4, 8 and 16 cores using two sockets.

By using Table \ref{perfDetail}, we can for instance analyze the performance and energy efficiency  of UPC, OpenMP and MPI for CG-4-2S and CG-8-2S presented in Figure \ref{fig:cgMops}. In CG-8-2S, MPI obtains the best performance and energy efficiency, UPC is second in performance and energy efficiency  and OpenMP is third. By using the data from Table \ref{perfDetail} it is possible to explain this result: UPC has an increased L2 cache hit ratio compared to OpenMP ($0.56~>~0.14$). MPI is better than OpenMP in CG-8-2S, because it has fewer L3 accesses ($11211~<~17023$) and better L2 hit ratio ($0.57~>~0.14$).

In MG-16-2S, UPC obtains the best performance compared to MPI and OpenMP because it has lower memory traffic than OpenMP ($452.36~GB~<~520.69~GB$) and MPI ($452.36~GB~<~487.4~GB$), better L3 cache hit ratio than OpenMP ($0.28~>~0.19$) and MPI ($0.28~>~0.2$) and fewer L3 accesses than OpenMP ($2782~<~3663$) and MPI ($2782~<~4390$).

In FT-16-2S, OpenMP obtains the best performance and energy efficiency  due to having lower memory traffic than UPC ($497.33~GB~<~1009~GB$) and MPI ($497.33~GB~<~878.8~GB$), better L3 hit ratio than UPC ($0.76~>~0.26$) and MPI ($0.76~>~0.43$) and a lower volume of L3 accesses than UPC ($1681~<~22562$) and MPI ($1681~<~8319$).

In IS-8-2S, MPI is better than OpenMP and UPC because it has higher L3 cache hit ratio than OpenMP ($0.13~>~0.04$) and UPC ($0.13~>~0.07$) and a lower level of L3 accesses than OpenMP ($2815~<~4550$) and UPC ($2815~<~5143$). UPC in IS-8-2S wins over OpenMP because of its slightly better L3 cache hit ratio ($0.07~>~0.04$).

Globally the results presented show a correlation between the number of cores used  and the achieved power efficiency. However we noticed in Table \ref{table:energyTable} a possible trade-off between performance and performance per watt by running programs over 1 or 2 sockets depending on what is the chosen goal (pure performance or energy efficiency).
We used Table \ref{perfDetail} to analyze the difference in performance between MPI, OpenMP and UPC over various threads/processes counts and sockets counts. We saw that by considering the memory traffic, L3 cache hit ratio, L2 cache hit ratio and the volume of access to L3 cache, that more insight into the performance can be obtained. 

\section{Conclusion}
\label{conclusion}

In this study, we have investigated and provided insights into UPC  energy efficiency and performance
using the latest CPU architecture with advanced support for energy and performance profiling. We have measured the energy efficiency and the computational performance of four kernels from the NAS Benchmark, both on a single-node system and on a multi-node supercomputer using three different programming models: UPC, MPI, and OpenMP.
On the multi-node supercomputer we observed that UPC is almost always inferior to MPI in terms of performance, although UPC scales well to 1024 cores and 64 nodes, the maximum system size used in this study.

From the measurements performed on the single-node system, we observed that by using more cores the performance and the energy efficiency both increase for the four selected kernels on the chosen hardware platform. The conclusion is not the same on the multi-node computer used in our experiments: the energy efficiency, except in one case, is not increasing with higher numbers of cores and nodes.

We would like to highlight that on the single-node system UPC can compete with MPI and OpenMP in terms of both computational speed and energy efficiency. We obtained this conclusion by analyzing the results produced on the single-node system in order to localize the origin of difference in performance.  We found that data locality is the main reason for the difference in performance. 

Our conclusions about UPC are compatible with results obtained in previous studies, in particular \cite{upcPerformance2009,shan2010programming}. We confirm that UPC can compete with both MPI and OpenMP in performance on a single-node.

In addition we provided a thorough analysis of the performance by looking at the L3 cache hit ratio, L2 cache hit ratio, memory traffic and L3 access of MPI, OpenMP and UPC. And we studied the results of UPC, MPI and OpenMP, running the selected kernels from the NAS Benchmarks on 2,4 and 8 cores by using one and two sockets in order to show the interest of the trade-off between energy efficiency and performance. 

In future work, we will explore hardware accelerators such as Many Integrated Cores (MIC) and GPUs. These  accelerators are well-known for being more energy efficient than CPUs for many applications.

Furthermore, it would be interesting to investigate why UPC, MPI and OpenMP differ in their communication pattern. In order to enhance the energy measurements, we aim for a more fine grained approach of measuring the energy consumption. By studying the energy cost of computation, communication between nodes, and between CPU and memory separately, we can suggest improvements to energy consumption both in the user codes and in the UPC compiler and runtime environment.

\vspace{-0.2cm}
\section*{Acknowledgments}
This work was supported by the European Union's Horizon 2020 research and innovation programme under grant agreement No.~671657, the European Union Seventh Framework Programme
(EXCESS project, grant n$^{\circ}$611183) and the Research Council of Norway (PREAPP project, grant n$^{\circ}$231746/F20).

\bibliographystyle{IEEEabrv}
\bibliography{paper}

\end{document}